\documentstyle[twoside,fleqn,espcrc2]{article}
\def\Z{$Z(3)$ }
\def\bn{{\bf n}}

\def\O{{\cal O}}
\def\Ot#1#2{\tilde{\cal O}^{#1}_{\bf #2}}
\def\Ht#1{{\tilde{\cal S}_{\bf #1}}}
\def\eval#1{\left\langle#1\right\rangle}

\newcommand{\AmS}{{\protect\the\textfont2
  A\kern-.1667em\lower.5ex\hbox{M}\kern-.125emS}}

\title{Effective spin models for the confinement phase transition\thanks{
Talk given at Lattice '98, the International Symposium on Lattice Field Theory, Boulder, Colorado, 13--18 July 1998}}

\author{Benjamin Svetitsky\address{School of Physics and Astronomy,
Raymond and Beverly Sackler Faculty of Exact Sciences,\\
Tel Aviv University,
69978 Tel Aviv, Israel}\thanks{Supported by the Israel Science Foundation
under Grant No.~255/96-1}}
       
\begin{document}

\begin{abstract}
Spatial correlations---bubbles, domain walls, etc.---can best be studied
by concentrating on the degrees of freedom most relevant to the
problem. For the finite temperature confinement transition, I integrate
out all gauge degrees of freedom, leaving only spins---Ising or
Potts---related to the Wilson line. I present problems that arise in the
course of this transformation and some results for the effective spin
action.
\end{abstract}

\maketitle

\section{EFFECTIVE MODELS}
The spatial structure of the
SU(3) gauge theory near the confinement phase
transition presents many interesting problems.
One might study the statics and dynamics of a planar boundary between the
high- and low-temperature phases 
\cite{surface} as well as the nucleation and
growth of bubbles and the possible stability of bubbles in either
phase \cite{bubbles}.
Straightforward lattice approaches to these properties are difficult
because of the large lattices needed.
We can make considerable progress by deriving an equivalent spin model
to study \cite{SW}.
Then, instead of working with
a non-Abelian gauge theory on a periodic four-dimensional
lattice of limited size, we can work on a three-dimensional
spin model in a much larger spatial volume.
The spin variables may also show domain structure at a glance that would
be hard to discern in the gauge theory.

Defining new degrees of freedom, let us associate $\sigma=+1$ with
one phase and $\sigma=-1$ with the other.
Then a general effective spin model will have the form
\begin{eqnarray}
S_{\rm spin}&=&\sum f(\bn-\bn')\sigma_{\bn}\sigma_{\bn'}+({\rm 3-spin})
\nonumber
\\&&\qquad+({\rm 4-spin})
+\cdots.
\end{eqnarray}
Simple restrictions of this action are the next-nearest-neighbor
Ising model
\begin{equation}
S=\beta\sum_{\rm nn}\sigma_\bn\sigma_{\bn'}
+\gamma\sum_{\rm nnn}\sigma_\bn\sigma_{\bn'}
\end{equation}
(recall the anisotropic version, the ANNNI model \cite{ANNNI})
and one of the simple models used to describe amphiphilic (oil--water--soap)
mixtures \cite{Widom},
\begin{eqnarray}
S&=&J\sum_{\rm nn}\sigma_\bn\sigma_{\bn'}
+h\sum\sigma_\bn+\gamma\sum_{\rm nnn}\sigma_\bn\sigma_{\bn'}\nonumber\\
&&\qquad+\delta\sum_{\rm triples}\sigma_\bn\sigma_{\bn'}\sigma_{\bn''}
\end{eqnarray}
When the couplings in these well-studied actions are chosen to give
competition between interactions, one finds 
a large variety of modulated equilibrium phases.
It would be exciting to discover such physics in our gauge theory,
but even if this doesn't happen, the advantages of the spin models
over the gauge theory are obvious.

\section{ISING ACTION }

Let's define the Ising variable $\sigma$ more precisely \cite{SW}.
We begin with the Wilson line,
\begin{equation}
L_\bn={\rm Tr}\,\prod_{\tau}U^0_{\bn,\tau}\ ,
\end{equation}
and define $\sigma_{\bn}=\{-1,+1\}$ according to whether $|L_\bn|$ is
less than or greater than some parameter $r$.
Refinements of this prescription might include smearing $L_\bn$ first,
or smearing and decimating (i.e., blocking).

Then we generate gauge configurations on an $N_t\times N_s^3$ lattice.
Each gauge configuration gives a spin configuration on an $N_s^3$
lattice.
From these configurations, we calculate an effective action for $\sigma$.
Choosing a set of operators $\O^{\alpha}$, we write a (truncated) action
\begin{equation}
S_{\rm eff}[\sigma]=\sum_{\alpha}\beta_{\alpha}\O^{\alpha}
\end{equation}
and evaluate the coefficients $\beta_{\alpha}$ via the Schwinger-Dyson
equations \cite{SDE} of the spin model,
\begin{equation}
\eval{\Ot{\alpha}n}=-\eval{\Ot{\alpha}n \exp{2\Ht n}}\ .
\end{equation}
($\Ot{\alpha}n$ is the piece of $\O^{\alpha}$ that contains the spin
$\sigma_{\bn}$, and similarly $\Ht n$.)

We expect that the effective couplings $\beta_{\alpha}$ will be continuous
functions of the gauge coupling $g$.
As $g$ passes through $g^*$, the gauge theory undergoes its confinement
phase transition.
We expect that $\beta_{\alpha}^*=\beta_{\alpha}(g^*)$ will be values of the
effective couplings at which the Ising model goes through a phase transition
from $\eval{\sigma}<0$ to $\eval{\sigma}>0$.

The result of the calculation \cite{SW} frustrates these expectations.
It turns out that the couplings $\beta_{\alpha}$ are {\em themselves\/} 
discontinuous at $g=g^*$ (which of course makes $\eval{\sigma}$ discontinuous
as expected).
Perhaps this isn't surprising:
The couplings $\beta_{\alpha}$ are derived from measured correlation functions
that are themselves discontinuous.
In any case,the situation
is reminiscent of a hoary controversy regarding the renormalization
group transformation near a first-order transition.
One school held that couplings in a blocked Hamiltonian must be continuous
functions of the unblocked couplings, and the transition is created by a
discontinuity fixed point.
The other school held that unblocked couplings on either side of the transition
will flow immediately to well-separated blocked couplings, creating the
discontinuity immediately.
The point was settled five years ago by theorems proved by van Enter,
Fern\'andez, and Sokal \cite{Sokal}, which state in brief that discontinuities in
the effective couplings (or the blocked couplings) are {\em impossible}.

To expand upon this a bit:
For each configuration, we map $U_{\bn,\tau}\to L_\bn \to \sigma_\bn$.
The original measure is $d\mu[U]=\exp{-S_W[U]}\,dU$, where $S_W$ is {\em local}.
We seek to replace this by a new measure $d\mu[\sigma]$.
At a first order transition, observables are discontinuous.
Nevertheless, say the theorems, $d\mu[\sigma]$ {\em must\/} turn out
continuous; otherwise $d\mu[\sigma]$ is non-Gibbsian, which means that
if it is to be written as $\exp{-S_{\rm eff}[\sigma]}\,d\sigma$, then
$S_{\rm eff}$ will not exist in the infinite volume limit.
The fault, of course, lies in the mapping.
We need better spin variables.

\begin{table*}[bt]
\setlength{\tabcolsep}{1.5pc}
\newlength{\digitwidth} \settowidth{\digitwidth}{\rm 0}
\catcode`?=\active \def?{\kern\digitwidth}
\caption{Terms in the effective 4-state Potts action and
couplings at the phase transition on a $2\times16^3$ lattice}
\label{table1}
\begin{tabular*}{\textwidth}{@{}l@{\extracolsep{\fill}}lr}
\hline
&\multicolumn{1}{c}{operator}&\multicolumn{1}{c}{coupling}\\
&\multicolumn{1}{c}{$\O^{\alpha}$}&\multicolumn{1}{c}{$\beta_{\alpha}^*$}\\
\hline
single-spin&
$\displaystyle \O^1=\vphantom{\sum_a^a}\sum_{\bf n}\delta(s_{\bf n},0)$&$-3.452$\\
nn1&
$\displaystyle \O^2=\vphantom{\sum_a^a}\sum_{\bf n}\sum_\mu 
\delta(s_{\bf n},s_{{\bf n}+\hat\mu})$&$-0.638$\\
nn2&
$\displaystyle \O^3=\vphantom{\sum_a^a}\sum_{\bf n}\sum_\mu 
\delta(s_{\bf n},0)\,\delta(s_{{\bf n}+\hat\mu},0)$&$0.862$\\
nnn1&
$\displaystyle \O^4=\vphantom{\sum_a^a}\sum_{\bf n}\sum_{\mu<\nu} 
\delta(s_{\bf n},s_{{\bf n}+\hat\mu\pm\hat\nu})$&$-0.104$\\
nnn2&
$\displaystyle \O^5=\vphantom{\sum_a^a}\sum_{\bf n}\sum_{\mu<\nu} 
\delta(s_{\bf n},0)\,\delta(s_{{\bf n}+\hat\mu\pm\hat\nu},0)$&$0.141$\\
$3^{\rm rd}\hbox{ neighbor}(1)\ \ $&
$\displaystyle \O^6=\vphantom{\sum_a^a}\sum_{\bf n}
\delta(s_{\bf n},s_{{\bf n}+\hat x \pm\hat y\pm\hat z})$&$-0.033$\\
$3^{\rm rd}\hbox{ neighbor}(2)$&
$\displaystyle \O^7=\vphantom{\sum_a^a}\sum_{\bf n}\delta(s_{\bf n},0)
\,\delta(s_{{\bf n}+\hat x \pm\hat y\pm\hat z},0)$&$0.046$\\
$4^{\rm th}\hbox{ neighbor}(1)$&
$\displaystyle \O^8=\vphantom{\sum_a^a}\sum_{\bf n}\sum_\mu 
\delta(s_{\bf n},s_{{\bf n}+2\hat\mu})$&$-0.049$\\
$4^{\rm th}\hbox{ neighbor}(2)$&
$\displaystyle \O^9=\vphantom{\sum_a^a}\sum_{\bf n}\sum_\mu \delta(s_{\bf n},0)
\,\delta(s_{{\bf n}+2\hat\mu},0)$&$0.064$\\
\hline
\end{tabular*}
\end{table*}

\section{POTTS ACTIONS}

Perhaps the definition of the Ising spin $\sigma$ may be modified to yield
a valid effective action.
The suspicion arises, however, that the simple projection used above
neglects some essential physics characteristic of the phase transition,
namely, the fact that it is a \Z order--disorder 
transition \cite{PRep}.

One may focus on the \Z physics by defining spins $P_\bn$ that encode
the phase of $L_\bn$, projected to the \Z directions.
This was done by Fukugita, Okawa, and Ukawa
\cite{FOU} who considered a 3-state Potts 
model with a general two-spin action,
\begin{equation}
S_{\rm Potts}=\sum f(|\bn-\bn'|)\,\delta(P_\bn ,P_\bn')
\end{equation}
The couplings $f(|\bn-\bn'|)$ were found to decay satisfyingly with
distance {\em and\/} to be continuous at the phase transition.
This action deserves to be studied further, in particular to see whether
it reproduces multi-spin correlations well.

The 3-state Potts variables do not, however, give a simple mapping of
ordered and disordered regions as do the Ising variables.
For this reason I have gone a step further, combining the Ising and
Potts degrees of freedom to give a {\em four\/}-state model.
I define $s_\bn=0$ if $|L_\bn|<r$, and $s_\bn=1,2,3$ according to the
phase of $L_\bn$ when $|L_\bn|>r$.
The effective action is that of a 4-state model with the $P_4$ symmetry
broken to $P_3$, truncated to a magnetic field term plus eight
two-spin couplings as shown in Table~\ref{table1}.
I show the couplings $\beta^*$ (preliminary, without error estimates)
at the phase transition for $N_t=2$, determined on
a $2\times16^3$ lattice without any of the smearing mentioned in connection
with the Ising action.
They are {\em continuous\/} across the transition.
We note that there is no apparent competition among the couplings.
Further study of this $S_{\rm eff}$ is in progress.

\end{document}